\documentclass[twocolumn]{article} 
\usepackage{geometry,comment}
\usepackage{amssymb}
\usepackage{amsmath}
\usepackage{multirow}
\usepackage{txfonts}
\usepackage{multicol}
\usepackage{booktabs}
\usepackage{mathdots}
\usepackage{graphicx} 
\usepackage{fancyhdr}
\usepackage{hyperref}
\usepackage{float}
\usepackage{microtype}
\usepackage{caption}
\usepackage{cite}
\RequirePackage{doi}
\usepackage{hyperref}

\usepackage{tikz}
\usepackage{mathrsfs}
\usetikzlibrary{arrows}
\usepackage{abstract}
\usepackage{blindtext}
\usepackage{placeins}
	
\usepackage{array}

\usepackage[switch,columnwise]{lineno}

\usepackage{sectsty}

\usepackage{tikz}
\usetikzlibrary{arrows}
\usepackage{csvsimple}
\usepackage{longtable}
\usepackage[T1]{fontenc}
\usepackage{authblk}
\makeatletter
\def\@xfootnote[#1]{%
  \protected@xdef\@thefnmark{#1}%
  \@footnotemark\@footnotetext}
\makeatother

\title{\textbf{Feature Selection on Lyme Disease Patient Survey Data}}

\author{Joshua Vendrow, Jamie Haddock, Deanna Needell, and Lorraine Johnson} 

\date{}

\definecolor{joshcolor}{rgb}{0.2, 0.8, 0.6}

\begin{document}

\twocolumn[
\maketitle

\begin{abstract} Lyme disease is a rapidly growing illness that remains poorly understood within the medical community.  Critical questions about when and why patients respond to treatment or stay ill, what kinds of treatments are effective, and even how to properly diagnose the disease remain largely unanswered.  We investigate these questions by applying machine learning techniques to a large scale Lyme disease patient registry, MyLymeData, developed by the nonprofit LymeDisease.org. We apply various machine learning methods in order to measure the effect of individual features in predicting participants' answers to the Global Rating of Change (GROC) survey questions that assess the self-reported degree to which their condition improved, worsened, or remained unchanged following antibiotic treatment. We use basic linear regression, support vector machines, neural networks, entropy-based decision tree models, and $k$-nearest neighbors approaches. We first analyze the general performance of the model and then identify the most important features for predicting participant answers to GROC. After we identify the ``key'' features, we separate them from the dataset and demonstrate the effectiveness of these features at identifying GROC. In doing so, we highlight possible directions for future study both mathematically and clinically. \end{abstract}\vspace{1cm}]

\section{Introduction} \label{introduction}

Lyme disease is the most common vector-borne disease in the United States. 
The CDC estimates that 300,000 people in the U.S. (approximately $1\%$ of the population) are diagnosed with Lyme Disease each year \cite{CDC13}, a rate 1.5 times higher than breast cancer \cite{bcancer16}, and six times higher than HIV/AIDS \cite{aids14}.

In its early, or acute, form, the disease may cause a hallmark erythema migrans (EM) rash and/or flu-like symptoms such as fever, malaise, fatigue, and generalized achiness \cite{aucott2009diagnostic}.  
A significant proportion of patients with Lyme disease develop chronic debilitating symptoms
that persist in the absence of initial treatment or following short-course antibiotic therapy \cite{aucott2013post}. This condition is commonly referred to as post-treatment Lyme disease or as chronic or persistent Lyme disease. In this paper, we refer to these patients as having persistent Lyme disease.

It is estimated that as many as $36\%$ of those diagnosed and treated early remain ill after treatment \cite{aucott2013post}. 
However, despite the high incidence and severity of Lyme disease, little research has been done, both clinically and analytically \cite{klempner2001two, krupp2003study, fallon2008randomized, delong2012antibiotic}. 
The result has been a stagnant and controversial research environment with little innovation and a costly lack of understanding or consensus. Physicians still do not know the best way to diagnose or treat Lyme, how it progresses, or why some patients respond to treatment and others do not.

{\bfseries Motivating questions.} 
We are motivated by questions that  interest both physicians and patients 
to better inform treatment approaches and to identify factors that might predict treatment response.

{\bf MyLymeData. }
Founded over 30 years ago, LymeDisease.org (LDo) is a national 501(c)(3) non-profit dedicated to advocacy, research and education. 
LDo has conducted surveys with the Lyme disease patient community since 2004, 
and published the results in peer reviewed journals.
In November 2015, LDo launched MyLymeData, a patient registry. MyLymeData has enrolled over 13,000 patients and continues to grow. Participants are asked hundreds of questions regarding their health history, diagnosis, symptoms, and treatment.

The first study using data from the registry was published in 2018. That study focused on treatment response variation among patients and identified a subgroup of high treatment responders using the Global Rating of Change Scale (GROC), a widely used and highly validated treatment response measurement \cite{johnson2018removing}. The GROC survey questions assess the degree to which 
participants reported that their condition improved, worsened, or remained unchanged following antibiotic treatment. We assign participants to class labels based on their GROC responder status. We label each participant as a high responder if they experienced substantial improvement, a low responder if they experienced slight improvement, and a non-responder if they worsened or remained unchanged. Medically, a major goal is to understand what patient attributes, protocols, or circumstances lead to patient improvement.  

{\bf Machine learning techniques. }
One challenge facing medical experts looking to derive insights from the data is that the high-dimensional structure of the data obscures the relationship between patient features and their GROC responder status.  In this work, we apply various machine learning models in order to measure the efficacy of individual features (survey question responses) in classifying the patients' GROC responder status, and to identify both meaningful and redundant information within the survey responses.  We apply both simple \emph{wrapper} and \emph{filter} approaches to feature selection \cite{guyon2003introduction}. \textit{We aggregate the results of several approaches to select a final subset of features we find most relevant to patients' GROC responder status, thereby highlighting what patient attributes and protocols are most likely associated with improved patient well being. These findings point to areas where additional analysis might prove useful.}

{\bf Organization. }
In Section \ref{Data_methods}, we describe the MyLymeData dataset and preprocessing steps, and introduce the techniques and models that we use throughout the paper. In Section \ref{general_accuracy}, we run all the models on the full MyLymeData dataset to evaluate the potential of each model to predict GROC labels using all of the data features. In Section \ref{identifying_key_features}, we apply the models to individual features to identify the features that are most important in predicting GROC labels and we form a subset of top features by aggregating the results of all the models. In Section \ref{restricting}, we evaluate the predictive ability of this subset of top features in comparison to the full dataset. Finally, in Section \ref{discussion} we discuss our results and their implication to Lyme Disease treatment protocols. 

\section{Data and Methods} \label{Data_methods}

Here we describe our experimental setup, the MyLymeData set, and the models and methods we use. 
\subsection{Experimental Setup} \label{setup}

All experiments are run on a MacBook Pro 2015 with a 2.5 GHz Intel Core i7 and a MacBook Pro 2018 with a 2.9 GHz Intel Core i9. We use Matlab version R2019b and Python version 3.7.3 to run experiments. We create linear regression models using Matlab's fitlm() function. We create neural networks models using Tensorflow; our network architecture is detailed in Section \ref{neuralnet}. We run support vector machine (SVM), Decision Tree, and $k$-nearest neighbors (KNN) models using the Python scikit-learn library. We run the SVM using the sklearn.svm.SVC() function, we run the Decision Tree with the sklearn.DecisionTreeClassifier() function and criterion=``entropy'' parameter, and we run the KNN model with the sklearn.KNeighborsClassifier() function. We use default hyperparameters for all functions, and experimentally choose optimal values of max depth for decision trees and K value for KNN. We directly calculate entropy in Section \ref{ind_entropy} for exact values. 

\subsection{The Dataset} \label{Dataset}

We use data from Phase 1 of the MyLymeData patient registry. Participants include respondents who report being US residents diagnosed with Lyme disease. 
We look only at participants who satisfy all of the following criteria:

\begin{enumerate}
\item Participant has persistent Lyme disease, which consists of patients who have experienced persistent symptoms for at least six months after antibiotic treatment. 
\item Participant responded that they were unwell.
\item Participant answered the GROC (Global Rating of Change) survey questions.
\end{enumerate}

We assign each participant a label based on their response to GROC, as previously described in \cite{johnson2018removing}. As asked, the GROC question produces a 17 point Likert scale. It is a two part question asking first if the patient is "better", "worse", or "unchanged". Patients who responded that they were better or worse are asked to specify the degree of improvement ranging from "almost the same" to "a very great deal better/worse". "Almost the same" responses for better or worse were combined with the unchanged response. As modified, the resultant 15 point Likert scale ranges between -7 and 7, with 0 as the midpoint for unchanged. We separate participants into three categories:

\begin{enumerate}
\item Non-responders, who answered between -7 and 0, indicating there was no improvement.
\item Low responders, who answered between 1 and 3, indicating there was slight improvement.
\item High responders, who answered between 4 and 7, indicating there was substantial improvement.

\end{enumerate}

Our dataset consists of 2162 participants who satisfy the necessary criteria and 215 features (question responses) drawn from the MyLymeData survey.  Each participant is assigned a label indicating non-responder, low responder, or high responder. The dataset has a total of 947 non-responders, 396 low responders, and 819 high responders. The 215 features cover diagnostic factors (such as delays to diagnosis, stage of diagnosis or presence of coinfections), treatment approach, individual antibiotic use and duration of use, alternative treatments, symptoms (severity, presence at time of diagnosis, and three worst), type of clinician, and degree of functional impairment. We refer to this dataset as MLD (MyLymeData), and we refer to all participants with a specific label as a class of participants.

In order to improve the survey-taking experience for participants, the format of the survey used a branching structure to direct only relevant questions to participants. Thus, for many of the features in our analysis only a subset of participants provided a response. For every such feature, we group all participants who did not respond to the feature together with a unique response. 

For our figures, we provide abbreviations of the relevant features. A name with a number following it indicates a series of questions within the same subtopic, and here we note such features with an ``i''. ``Bio\_Sex'' indicates biological sex. ``Sx\_Dx\_i'' indicate symptoms present at diagnosis. ``Tick'' indicates the presence of a tick bite. ``Sx\_Sev\_i'' indicate the severity of specific symptoms. ``Sx\_Top\_i'' indicate a specific symptom as being in the top 3 symptoms. ``Abx'' indicates whether the participant is currently taking antibiotics and/or alternative treatments. ``Abx\_Not\_i'' indicate the reasons that a participant is current not taking antibiotics. ``Abx\_Dur'' indicates the duration of the current antibiotic treatment protocol. ``Abx\_Eff'' indicates the effectiveness of the current antibiotic treatment protocal. ``Abx\_Oral,'' ``Abx\_IM,'' and ``Abx\_IV'' indicate whether the current antibiotic protocal includes oral, IM, and/or IV antibiotics, respectively. ``Abx\_i'' indicate whether the participant is currently taking a specific oral antibiotic. ``Abx\_IM\_i'' indicate whether the participant is current taking a specific IM antibiotic. ``Alt\_Tx\_Eff\_i'' and ``Alt\_Tx\_Sfx\_i'' indicate the effectiveness of and side effects of current alternative treatment approaches, respectively. ``Med\_i'' indicate whether a participant is taking a specific non-antibiotic medication. ``Provider\_i'' indicate whether a participant's Lyme disease is being treated by a specific type of healthcare provider. ``Wrk\_Day'' indicates the number of times a participant went to work but was unable to fully concentrate because of not feeling well. These features are more fully described in Table \ref{table:attachment} of the Appendix.

\subsection{Techniques}

Here we introduce the methods that we utilize through our experiments to help us assess importance of the features in determining participants' GROC responder status.

\subsubsection{Subsampling} \label{subsample}

Most of the machine learning techniques that we use require balanced class sizes to produce accurate results. For this reason, we subsample the the data by selecting participants from MLD so that there are an even number of non-responders, low responders, and high responders. Thus, this dataset has 396 participants with each label. We refer to the subsampled dataset as SMLD (subsampled MyLymeData). We will use a fixed subsample throughout the experiments for consistency. 

\subsubsection{Permuting Labels} \label{permute}

We wish to show that the models created from our datasets outperform the same models on similar, but random and meaningless data. This is important because some models overfit even to random data given an appropriate number of datapoints and features.
We compare our results on MLD to the results on similar but random data.
We create such a dataset by randomly permuting the labels and reassociating them with participants. We perform this on both MLD and SMLD and call these random datasets RMLD (random MyLymeData) and RSMLD (random subsampled MyLymeData).

\subsubsection{Dropping Features} \label{Dropping}

One way we assess the contribution of each feature to the model's predictive ability is by evaluating the accuracy of a model with the entire dataset, and then removing a feature from the dataset and reevaluating the accuracy. Ideally, the most important features would cause the largest decrease in accuracy when they are removed from the dataset. This metric allows us to assess how influential each feature is to the model.

\subsubsection{Validation Function} \label{Validation}

For models that are prone to overfitting, we use a function that effectively evaluates validation accuracy. This function accepts a dataset and a classifier, and over a specified amount of trials (30 by default), creates a random 75/25 training and testing split of the data that is different on every trial, fits the classifier to the training data, and measures accuracy on the testing data. We will call this function our validation function. 

\subsection{Models}

Our machine learning methods are all supervised learning methods, meaning that the model aims to learn a mapping from an input, in this case our participants and their responses, to an output, in this case the three classes we identified. We use two distinct supervised learning methods, regression and classification. Regression models produce a continuous value that numerically approximates the output, while classification models produce a discrete output, in this case one of the three classes. In our classification models, we calculate accuracy A as 
\begin{equation}
\label{eq:1}
A = \dfrac{|T_c|}{|T|}
\end{equation}
where $|T|$ denotes the total amount of participants and $|T_c|$ denotes the total amount of participants that were predicted correctly.

\subsubsection{Linear Regression}

Linear regression is a regression model that attempts to produce the best affine hyperplane to fit a dataset.  The model computes the optimal affine hyperplane that minimizes the sum of the squares of the distances of the points from their projections onto the hyperplane along the dependent variable coordinate subspaces. 

\subsubsection{SVM}

A support vector machine (SVM) is a popular and widely used classification model in the area of machine learning. The model attempts to separate datapoints of different classes with an affine hyperplane. The SVM aims to find the optimal hyperplane by reducing both the amount of points classified incorrectly and the distance of these incorrectly classified points from the hyperplane. To achieve our multiclass classification task, SVM separates the three classes via three hyperplanes that attempt to separate each class from the other classes. In our experiments using the SVM model when dropping out features (see Section \ref{Dropping}), we examine the training accuracy to measure the separability of the dataset by an affine hyperplane.

\subsubsection{Neural Network} \label{neuralnet}

We train a neural network model with two dense layers. In Figure \ref{fig:NNarchitecture}, we display the architecture of our neural network. Each hidden layer has 20 nodes. We use a softmax output layer for multiclass classification. We compile our model with an Adam optimizer \cite{kingma2014adam}.

\newcommand\layersep{2.1cm}
\newcommand\layerseps{4.2cm}
\newcommand\width{7cm}
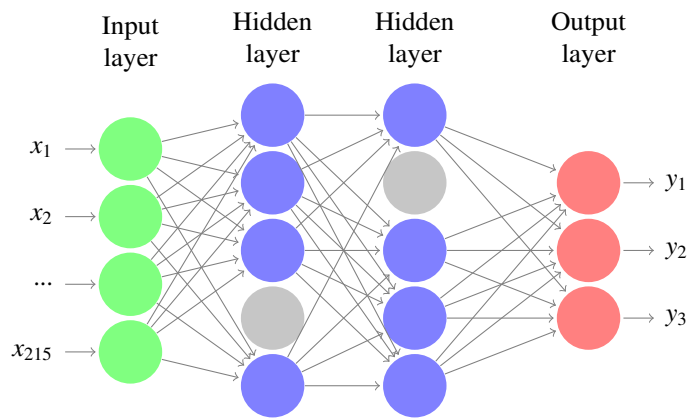
\begin{figure}[H]
\begin{tikzpicture}[shorten >=1pt,->,draw=black!50, node distance=\layerseps,scale=0.9]
\tikzstyle{every pin edge}=[<-,shorten <=1pt]
\tikzstyle{neuron}=[circle,fill=black!25,minimum size=24pt,inner sep=1pt]
\tikzstyle{input neuron}=[neuron, fill=green!50];
\tikzstyle{output neuron}=[neuron, fill=red!50];
\tikzstyle{hidden neuron}=[neuron, fill=blue!50];
\tikzstyle{hidden neuron2}=[neuron, fill=blue!50];
\tikzstyle{drop}=[neuron, fill={rgb:black,2;white,7}];
\tikzstyle{annot} = [text width=4em, text centered]

\foreach \name / \y in {1,2}
\node[input neuron, pin=left:$x_{\name}$] (I-\name) at (0,-\y) {};

\node[input neuron, pin=left:...] (I-3) at (0,-3) {};
\node[input neuron, pin=left:$x_{215}$] (I-4) at (0,-4) {};

\foreach \name / \y in {1,...,3}
\path[yshift=0.5cm]
node[hidden neuron] (H-\name) at (\layersep,-\y cm) {};

\foreach \name / \y in {5}
\path[yshift=0.5cm]
node[hidden neuron] (H-\name) at (\layersep,-\y cm) {};

\path[yshift=0.5cm]
node[drop] (H-4) at (\layersep,-4 cm) {};

\foreach \name / \y in {1}
\path[yshift=0.5cm]
node[hidden neuron2] (H2-\name) at (\layerseps,-\y cm) {};

\foreach \name / \y in {3,...,5}
\path[yshift=0.5cm]
node[hidden neuron2] (H2-\name) at (\layerseps,-\y cm) {};

\path[yshift=0.5cm]
node[drop] (H2-2) at (\layerseps,-2 cm) {};

\foreach \y [count = \name] in {2,3,4}
\node [output neuron,pin={[pin edge={->}]right:$y_{\name}$}, right of=H-\y] (J-\name) {};

\foreach \source in {1,...,4}
\foreach \dest in {1,...,3}
\path (I-\source) edge (H-\dest);

\foreach \source in {1,...,4}
\foreach \dest in {5}
\path (I-\source) edge (H-\dest);

\foreach \source in {1,2,3,5}
\foreach \dest in {1,3,4,5}
\path (H-\source) edge (H2-\dest);

\foreach \source in {1}
\foreach \dest in {1,2,3}
\path (H2-\source) edge (J-\dest);

\foreach \source in {3,...,5}
\foreach \dest in {1,2,3}
\path (H2-\source) edge (J-\dest);

\node[annot,above of=H-1, node distance=1cm] (hl) {Hidden layer};
\node[annot,above of=H2-1, node distance=1cm] (h2) {Hidden layer};
\node[annot,above of=I-1,node distance=1.4cm] {Input layer};
\node[annot,above of=J-1,node distance=1.9cm] {Output layer};
\end{tikzpicture}
    \caption{Neural Network Architecture}
    \label{fig:NNarchitecture}
\end{figure}

We train each model for 50 epochs, train 30 models, and of these 30 models choose the model with the highest validation accuracy. The neural network objective function is nonconvex so we do multiple runs to avoid poor local minima. We then measure the accuracy of our best model using additional test data. We repeat this entire process multiple times and average over the results to attain an average test accuracy which we use to evaluate the effectiveness of the model. This process is more complex than our other models, but is valuable because of the nonlinearity and expressive capability of the model. We also increase the train/test ratio to 0.8/0.2 for our neural network to allow for sufficient training data. 

Due to computational constraints, we do not run the neural network when selecting top features, but only to evaluate general performance and evaluate the effectiveness of the top features we select. Because the neural network objective is nonconvex, there is variation in each run so we would need to run the neural network many times to be able to find significant differences between features which is computationally prohibitive.

\subsubsection{Entropy and Decision Tree}

In order to create the decision tree, we use the entropy metric to measure the importance of each feature. The goal of entropy is to calculate the randomness of the data, so we measure feature importance by calculating the decrease in entropy after the data is split by the feature. Let X be a discrete random variable that takes on a value of one of the three labels with probability
\begin{equation}
\label{eq:1}
p_X(i) = \dfrac{|T_i|}{|T|}
\end{equation}
where $|T|$ denotes the total participants and $|T_i|$ denotes the participants with label i. Then, We measure entropy as
\begin{equation}
\label{eq:1}
H[X] = -\mathbb{E}(\log(X)) = -\displaystyle\sum_i p_X(i)\log(p_X(i))
\end{equation} 

\vspace{5pt}
Using this criterion, we measure the importance of a feature by comparing the entropy of the dataset to the conditional entropy of the dataset after the dataset is split based on the participants' responses to this question. We refer to the decrease in entropy as information gain. 

We create a decision tree that assesses feature importance using the entropy criterion. To create this tree, the scikit-learn model places the most important features highest in the tree to improve its ability to split the data based on class labels, so at every node the function uses the feature that most effectively decreases entropy. In order to prevent overfitting with the decision tree, we run our tree model on the validation function as outlined in section 1.2.4, once at every depth, and choose the depth that produces the maximum validation accuracy. 

\subsubsection{$k$-Nearest Neighbors}

The KNN algorithm classifies an example by looking at the $k$ points nearest to it and selecting the most common label amongst these points. We measure the distance between points using Euclidean distance. The most important decision when training this model is the choice of $k$. We split our data into training data and validation data, and we choose K by training our model on the training data with various choices of $k$ and choosing the value that produces the maximum accuracy by the validation function as outlined in Section \ref{Validation}.  

\section {Results}  \label{results}

First, we run each of our models on the complete dataset in order to evaluate the potential of each model to predict GROC labels using all of the data features. We then run our models on individual features in order to identify features that are important in predicting GROC labels, and we aggregate these results into a subset of ``key" features. Finally, we run our models on the two subsets of only the key features and only the remaining features to measure the effectiveness of our identified key feature set at predicting GROC labels.

\subsection{General Performance} \label{general_accuracy}

We first run each of our models on our datasets to evaluate the potential of each model to predict GROC labels using all of the data features.  This gives us a measure to compare against in determining the importance of each feature in the next section. We also 
demonstrate the inaccuracy of the model at predicting undersampled labels in MLD which motivates subsampling the data.

In Table \ref{table:general_acc} we list the results of running each model on four datasets: MLD, randomized MLD (RMLD), subsampled MLD (SMLD), and randomized subsampled MLD (RSMLD) (details for the construction of these datasets is provided in Sections \ref{subsample} and \ref{permute}). We list prediction accuracies for our classification models, and relevant regression values for the linear regression model.

\begin{table}[H]
\centering
\caption{Results for running each model on MLD, randomized MLD (RMLD), subsampled MLD (SMLD), and random subsampled MLD (RMLD). For linear regression we list regression values and for our classificaiton models we list accuracies measured by the validation function (details in \ref{Validation}).}
\begin{tabular}{l l l l l}
  \hline \bfseries{Model} & \bfseries{MLD} & \bfseries{RMLD}  & \bfseries{SMLD}  & \bfseries{RSMLD}\\\hline
  
  SVM & 0.576 & 0.400 & 0.484 & 0.342 \\
  Neural Net & 0.598 & 0.422 & 0.518 & 0.325 \\
  Decision Tree & 0.600 & 0.526 & 0.432 & 0.335 \\
  KNN & 0.542 & 0.439 & 0.490 & 0.335 \\
\\
   Linear Regression \\
   \quad $R^2$ & 0.397 & 0.095 & 0.416 & 0.175 \\
   \quad Adjusted $R^2$ & 0.331 & -0.004 & 0.290 & -0.003 \\
   \quad p-value & 3.7E-108 & 0.439 & 6.2E-36 & 0.400 \\
   \quad RMSE & 0.738 & 0.904 & 0.688 & 0.818 \\
   \hline
\end{tabular}
\label{table:general_acc}
\end{table}

We see that for classification on randomized MLD (RMLD), we achieve above-random accuracies (significantly larger than 0.333), which can be attributed to a a variation in class sizes, while the classification accuracies for randomized subsampled MLD (RSMLD) are all approximately 0.333. This suggests that a part of the high accuracies reported for MLD in comparison to SMLD can be attributed to factors unrelated to the predictive ability of the model. 

In Tables \ref{table:mld_groc} and \ref{table:smld_groc}, we display the accuracy of the MLD and SMLD datasets at prediction for participants of each GROC class. We see that for all of our models, the classification accuracies for MLD vary significantly across GROC class, and that the low responders, the class with the least participants, has very poor prediction accuracy. For each of our models, this variation is significantly reduced for classification on SMLD.

\begin{table}[H]
\centering
\caption{Prediction accuracies of each classification model on  MLD for participants of each GROC class (non-responders, low responders, high responders).}
\begin{tabular}{l l l l}
  \hline \bfseries{Model} & \bfseries{Non} & \bfseries{Low}  & \bfseries{High}\\\hline
  
  SVM & 0.709 & 0.193 & 0.605 \\
  Neural Net & 0.728 & 0.139 & 0.635 \\
  Decision Tree & 0.776 & 0.107 & 0.636 \\
  KNN & 0.691 & 0.033 & 0.613 \\

   \hline
\end{tabular}
\label{table:mld_groc}
\end{table}

\begin{table}[H]
\centering
\caption{Prediction accuracies of each classification model on subsampled MLD (SMLD) for participants of each GROC class (non-responders, low responders, high responders).}
\begin{tabular}{l l l l}
  \hline \bfseries{Model} & \bfseries{Non} & \bfseries{Low}  & \bfseries{High}\\\hline
  
  SVM & 0.546 & 0.383 & 0.522 \\
  Neural Net & 0.487 & 0.465 & 0.606 \\
  Decision Tree & 0.641 & 0.305 & 0.634 \\
  KNN & 0.509 & 0.429 & 0.534 \\

   \hline
\end{tabular}
\label{table:smld_groc}
\end{table}

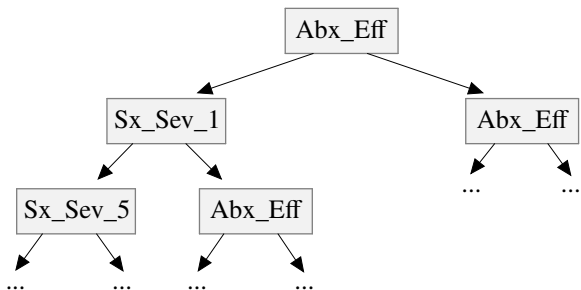
\begin{figure}[H]
    \centering
    \definecolor{yqyqyq}{rgb}{0.5019607843137255,0.5019607843137255,0.5019607843137255}
\begin{tikzpicture}[line cap=round,line join=round,>=triangle 45,x=2.0cm,y=2.0cm,scale=0.3]
\clip(-6.5,-1.42) rectangle (6.54,5.16);
\fill[line width=1.pt,color=yqyqyq,fill=yqyqyq,fill opacity=0.10000000149011612] (-0.2,5.) -- (-0.2,4.) -- (2.3,4.) -- (2.3,5.) -- cycle;
\fill[line width=1.pt,color=yqyqyq,fill=yqyqyq,fill opacity=0.10000000149011612] (3.8,3.) -- (6.3,3.) -- (6.3,2.) -- (3.8,2.) -- cycle;
\fill[line width=1.pt,color=yqyqyq,fill=yqyqyq,fill opacity=0.10000000149011612] (-4.15,3.) -- (-1.52,3.) -- (-1.52,2.) -- (-4.15,2.) -- cycle;
\fill[line width=1.pt,color=yqyqyq,fill=yqyqyq,fill opacity=0.10000000149011612] (-2.1,1.) -- (0.4,1.) -- (0.4,0.) -- (-2.1,0.) -- cycle;
\fill[line width=1.pt,color=yqyqyq,fill=yqyqyq,fill opacity=0.10000000149011612] (-3.52,1.) -- (-3.52,0.) -- (-6.15,0.) -- (-6.15,1.) -- cycle;
\draw [line width=0.5pt,color=yqyqyq] (-0.2,5.)-- (-0.2,4.);
\draw [line width=.5pt,color=yqyqyq] (-0.2,4.)-- (2.3,4.);
\draw [line width=.5pt,color=yqyqyq] (2.3,4.)-- (2.3,5.);
\draw [line width=.5pt,color=yqyqyq] (2.3,5.)-- (-0.2,5.);
\draw [line width=.5pt,color=yqyqyq] (3.8,3.)-- (6.3,3.);
\draw [line width=.5pt,color=yqyqyq] (6.3,3.)-- (6.3,2.);
\draw [line width=.5pt,color=yqyqyq] (6.3,2.)-- (3.8,2.);
\draw [line width=.5pt,color=yqyqyq] (3.8,2.)-- (3.8,3.);
\draw [line width=.5pt,color=yqyqyq] (-4.15,3.)-- (-1.52,3.);
\draw [line width=.5pt,color=yqyqyq] (-1.52,3.)-- (-1.52,2.);
\draw [line width=.5pt,color=yqyqyq] (-1.52,2.)-- (-4.15,2.);
\draw [line width=.5pt,color=yqyqyq] (-4.15,2.)-- (-4.15,3.);
\draw [line width=.5pt,color=yqyqyq] (-2.1,1.)-- (0.4,1.);
\draw [line width=.5pt,color=yqyqyq] (0.4,1.)-- (0.4,0.);
\draw [line width=.5pt,color=yqyqyq] (0.4,0.)-- (-2.1,0.);
\draw [line width=.5pt,color=yqyqyq] (-2.1,0.)-- (-2.1,1.);
\draw [line width=.5pt,color=yqyqyq] (-3.52,1.)-- (-3.52,0.);
\draw [line width=.5pt,color=yqyqyq] (-3.52,0.)-- (-6.15,0.);
\draw [line width=.5pt,color=yqyqyq] (-6.15,0.)-- (-6.15,1.);
\draw [line width=.5pt,color=yqyqyq] (-6.15,1.)-- (-3.52,1.);
\draw [->,line width=.3pt] (0.42,4.) -- (-2.16,3.12);
\draw [->,line width=.3pt] (1.62,4.) -- (4.2,3.12);
\draw [->,line width=.3pt] (-3.58,2.) -- (-4.36,1.2);
\draw [->,line width=.3pt] (-2.4,2.) -- (-1.6,1.2);
\draw [->,line width=.3pt] (4.44,2.) -- (3.9,1.32);
\draw [->,line width=.3pt] (5.62,2.) -- (6.14,1.28);
\draw [->,line width=.3pt] (-5.58,0.) -- (-6.18,-0.82);
\draw [->,line width=.3pt] (-4.38,0.) -- (-3.78,-0.84);
\draw [->,line width=.3pt] (-1.56,0.) -- (-2.14,-0.84);
\draw [->,line width=.3pt] (-0.36,0.) -- (0.22,-0.86);
\draw (-0.18,4.94) node[anchor=north west] {Abx\_Eff};
\draw (3.84,2.96) node[anchor=north west] {Abx\_Eff};
\draw (-4.20,2.94) node[anchor=north west] {Sx\_Sev\_1};
\draw (-2.07,0.96) node[anchor=north west] {Abx\_Eff};
\draw (-6.2,0.96) node[anchor=north west] {Sx\_Sev\_5};
\draw (3.5,1.2) node[anchor=north west] {$...$};
\draw (5.7,1.2) node[anchor=north west] {$...$};
\draw (-6.6,-0.96) node[anchor=north west] {$...$};
\draw (-4.24,-0.96) node[anchor=north west] {$...$};
\draw (-2.58,-0.96) node[anchor=north west] {$...$};
\draw (-0.2,-0.96) node[anchor=north west] {$...$};
\end{tikzpicture}
    \caption{Decision tree built by the scikit-learn library for SMLD using the entropy metric. The ellipses indicate the continuation of the tree for each direction. At each step, the model selects the feature that causes the largest decrease in entropy for the current subset of the full data to split the tree. Note that each split along the tree is a binary split, so we can have multiple splits along Abx\_Eff because it has more than two possible responses.}
    \label{fig:decision_tree}
\end{figure}

These preliminary results suggest to us that evenly subsampling our MLD dataset could lead to more accurate and meaningful results despite the higher classification accuracy for MLD. For this reason, for the remainder of our classification tasks we use SMLD.  
These regression and classification results also suggest to us that there is a substantial relationship between a participant's GROC class and certain survey responses. Our goal in the next section will be to identify the specific features with the most significant relationship to GROC class.

\subsection{Identifying Key Features} \label{identifying_key_features}
Here, we run our models on individual features in order to identify the features that are most important in predicting GROC labels. We list notable results here and will highlight those features that are ranked highly by several models.

\subsubsection{Linear Regression}

\begin{figure}[H]
  \centering
      \includegraphics[width=0.9\columnwidth]{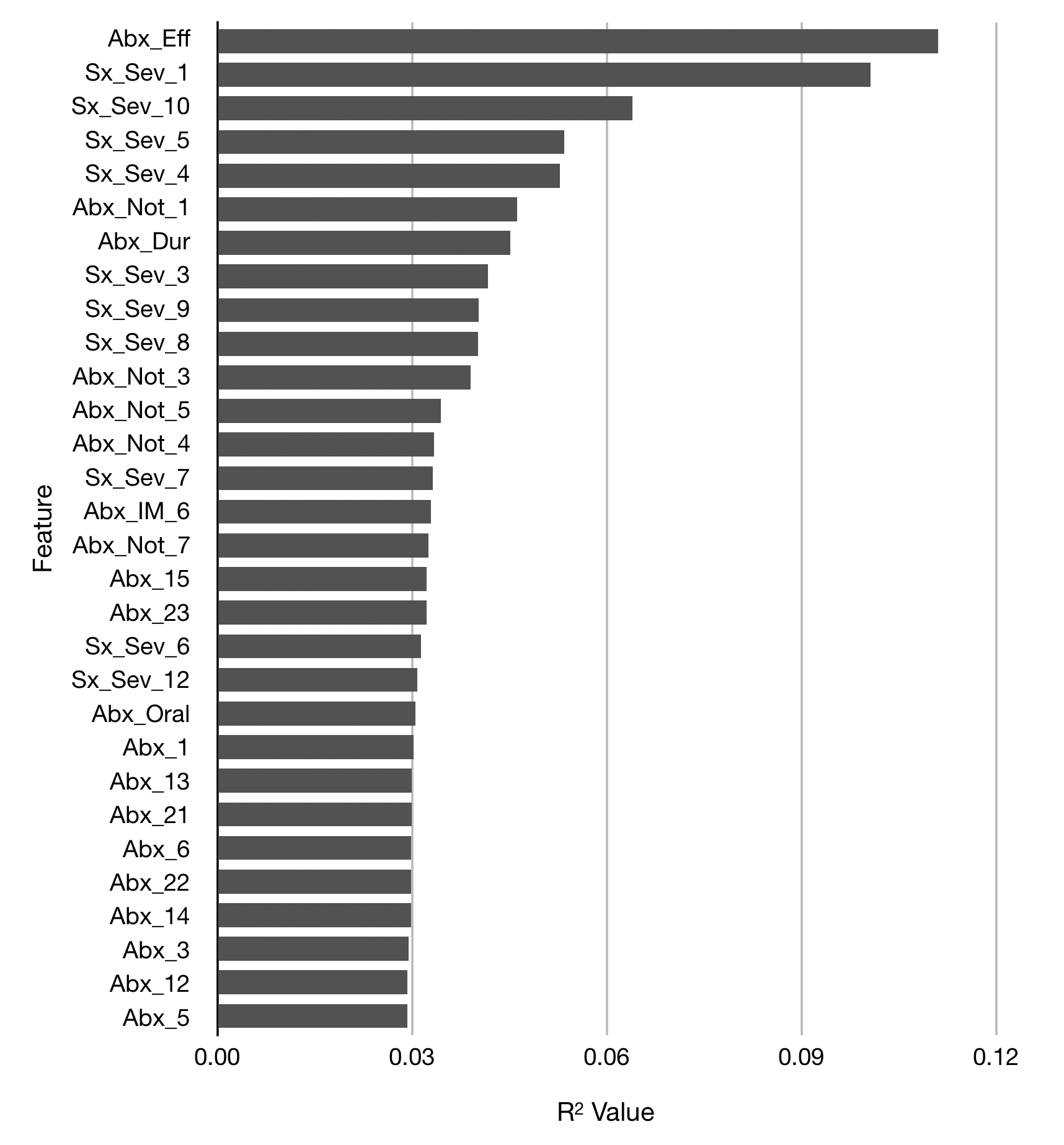}

        \caption{
                \label{fig:exp_plots}  
                $R^2$ values of individual features on Subsampled MLD (SMLD). Here we display the top 30 features by $R^2$ value in descending order.
        }
        \label{fig:rindividual}
\end{figure}

In Figure \ref{fig:rindividual} we show the results of running linear regression on SMLD and measuring the $R^2$ values of the individual features; see Section \ref{setup} for experimental design details. Here, we calculate $R^2$ values by running a separate single-variable regression for each feature.

\subsubsection{SVM}

\begin{figure}[H]
  \centering
      \includegraphics[width=0.9\columnwidth]{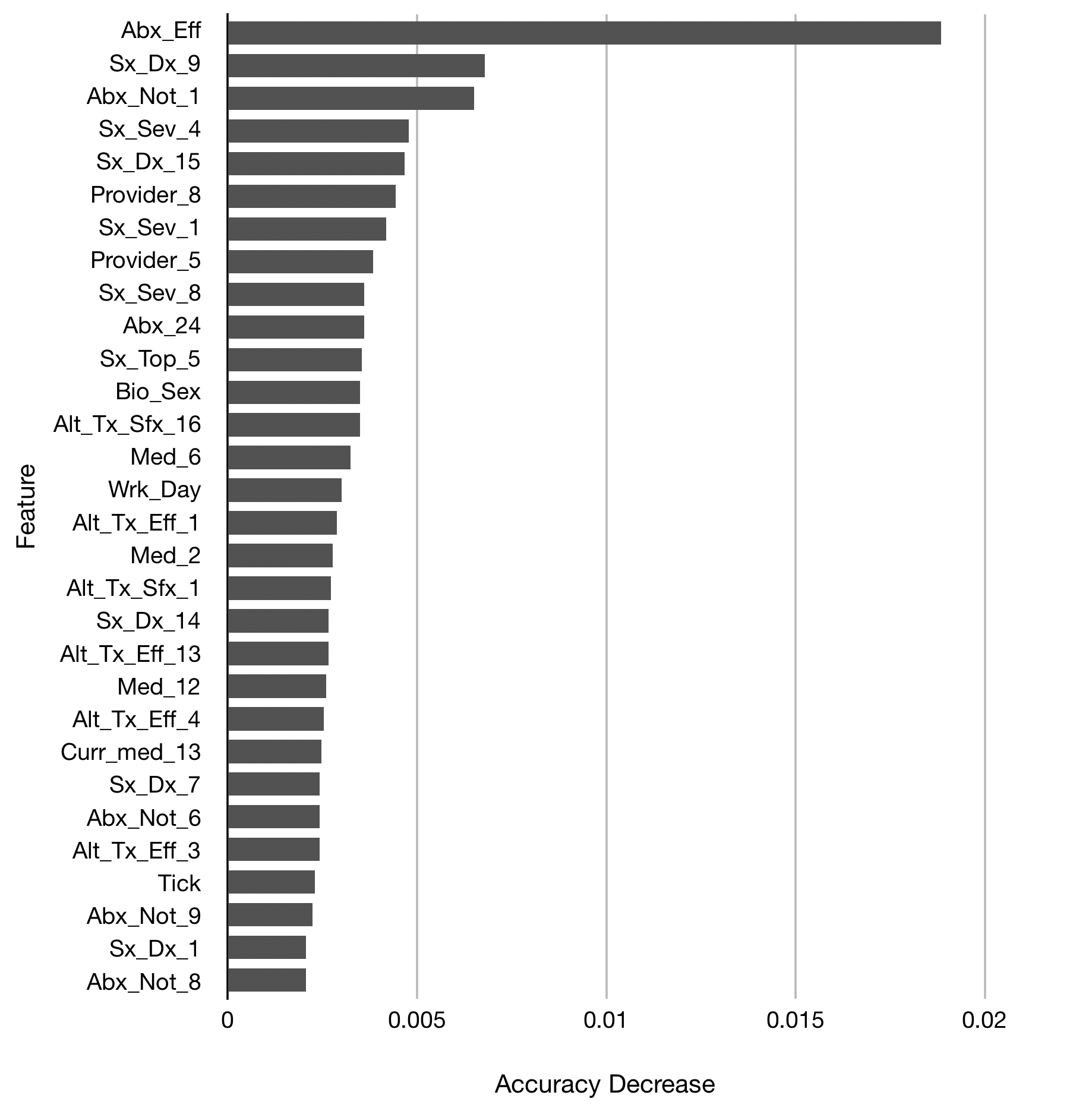}

        \caption{
                \label{fig:exp_plots}  
                SVM training accuracy dropping individual deatures on Subsampled MLD (SMLD). For each feature, we measure the decrease in accuracy caused by removing a single feature from the dataset. Here we display the top 30 features by accuracy in descending order.
        }
        \label{fig:svmindividual}
\end{figure}

In Figure \ref{fig:svmindividual} we record the change in training accuracy of the SVM on SMLD after dropping each feature from the data set individually; see Section \ref{setup} for experimental design details and see Section \ref{Dropping} for details about the dropout process. In order to accurately assess the effect of dropping each feature, we take random subsets of SMLD and run the model on each subset of the data. Using this method, we suggest that the magnitude of decrease in accuracy when dropping a feature demonstrates its ability to predict GROC value. For this reason, in Figure \ref{fig:svmindividual} we also sort the features in order of decreasing accuracy drop. 

One weakness of this metric is that small impacts to the shape of the dataset from dropping out a feature can cause random changes to testing accuracy that make our ranking less precise. To address this we use training accuracy to measure the separability of the dataset.

\begin{figure}[H]
  \centering
      \includegraphics[width=0.9\columnwidth]{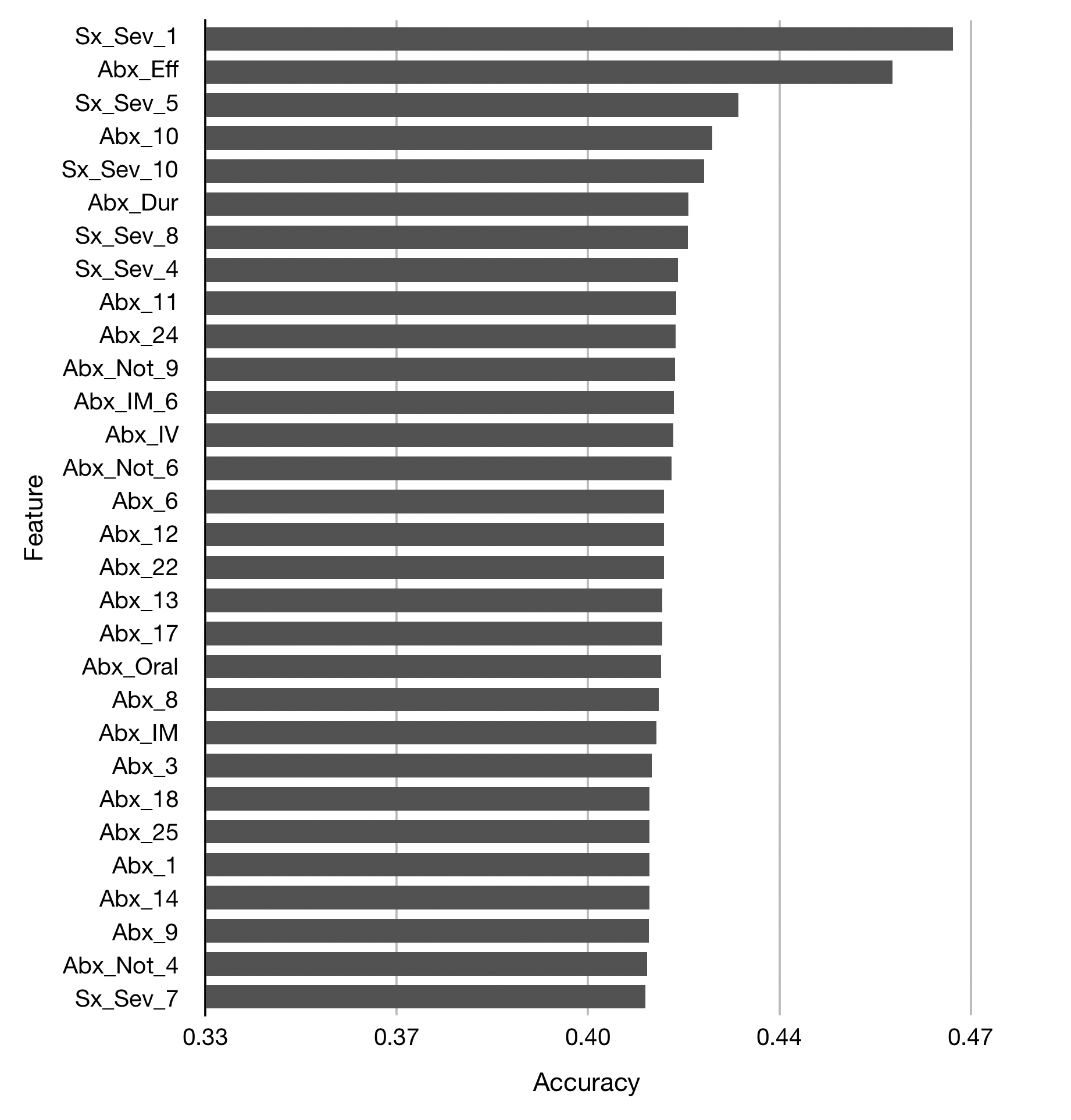}

        \caption{
                \label{fig:exp_plots}  
                SVM accuracy of individual featues on Subsampled MLD (SMLD). We attain this accuracy by running our SVM model on only the single feature. Here we display the top 30 features by accuracy in descending order. We measure accuracies by the validation function (details in \ref{Validation}).
        }
        \label{fig:svmsingleindividual}
\end{figure}

In Figure \ref{fig:svmsingleindividual} we show the results of running the SVM model on only individual features rather than dropping the feature from the entire dataset. 

\subsubsection{Entropy / Decision Tree} \label{ind_entropy}

\begin{figure}[H]
  \centering
      \includegraphics[width=0.9\columnwidth]{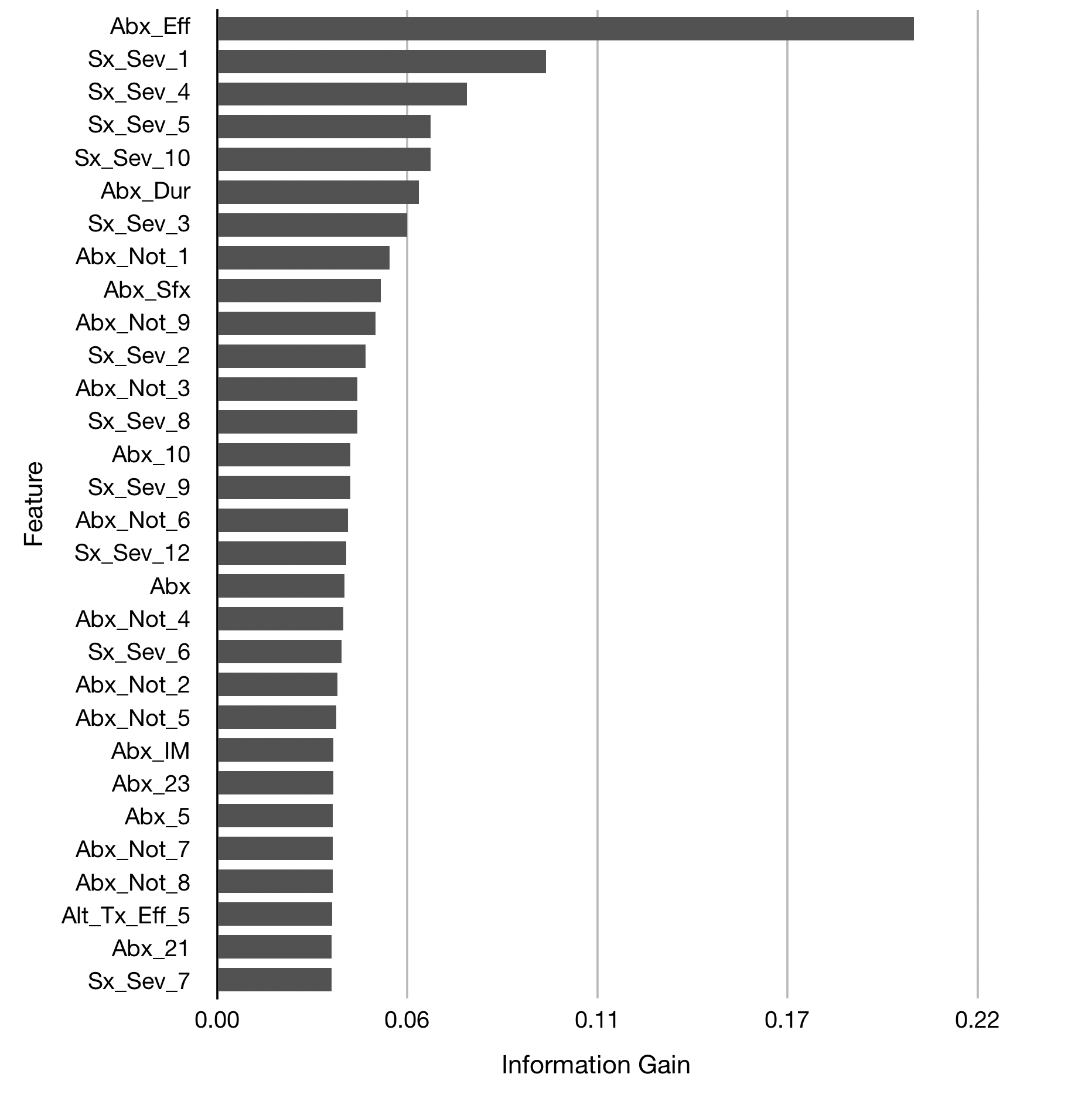}

        \caption{
                \label{fig:exp_plots}  
                Information gain of individual features on MLD. Here we display the top 30 features by entropy in descending order. We measure accuracies by the validation function (details in \ref{Validation}).
        }
        \label{fig:entropyindividual}
\end{figure}

We measure the information gain produced by each feature by calculating the difference of the original entropy of the features and the conditional entropy after splitting the data into separate groups based on the responses to this feature. For entropy calculations, we do not use the subsampled data because the difference in class sizes does not negatively affect performance as we are not performing a classification task that could result in poor classification accuracy on underrepresented labels. 
Meanwhile, the decision tree model, which uses this entropy calculation for classification, does suffer from uneven prediction results (e.g., a one node decision tree degenerates to a majority classifier) so for classification we apply the model to the subsampled dataset.
In Figure \ref{fig:entropyindividual} we list the top entropy gain produced by individual features in MLD, sorted in descending order of entropy gain. 

\subsubsection{$k$-Nearest Neighbors}

\begin{figure}[H]
  \centering
      \includegraphics[width=0.9\columnwidth]{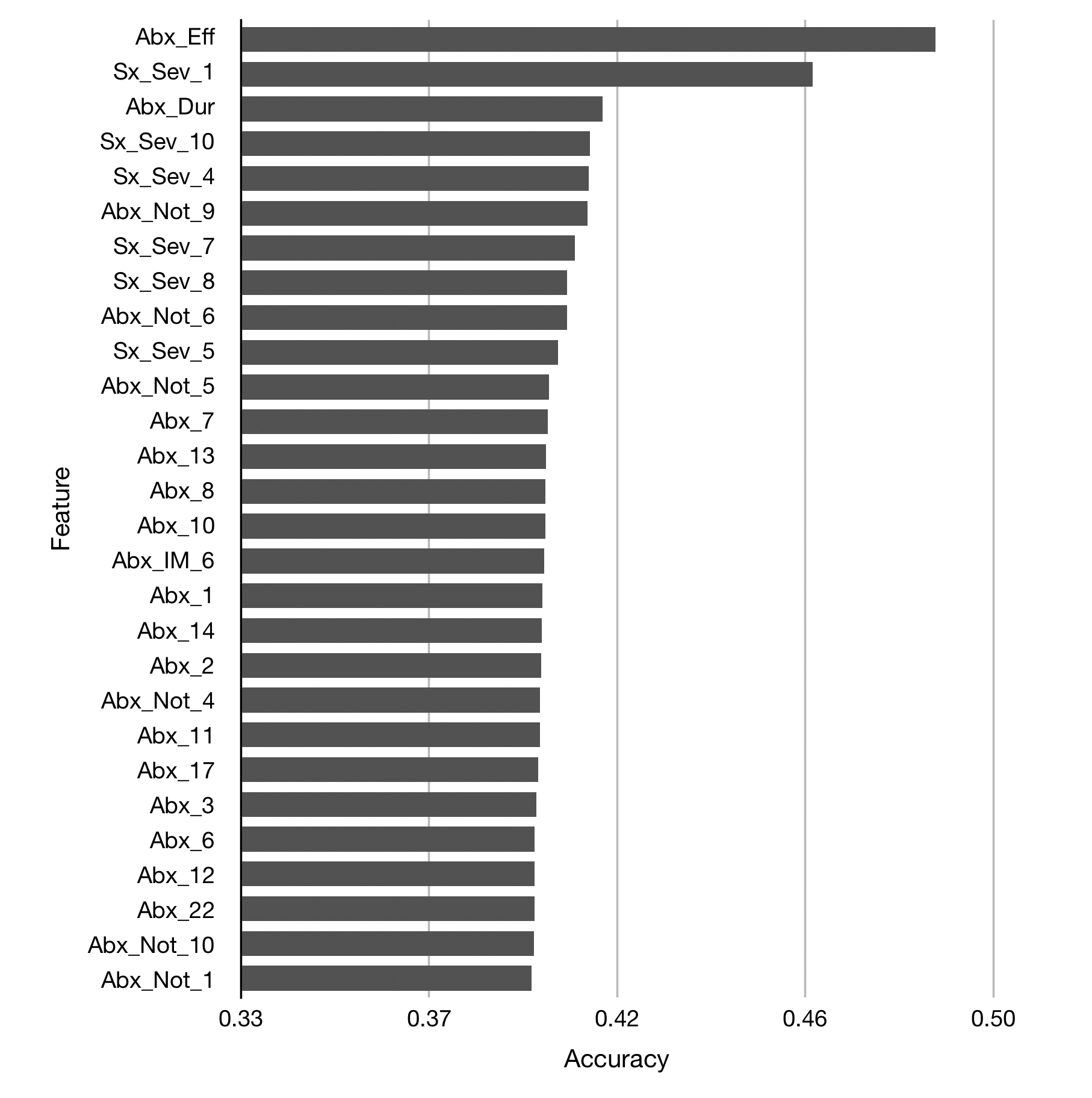}

        \caption{
                \label{fig:exp_plots}  
                KNN accuracy of individual features on Subsampled MLD (SMLD). Here we display the top 30 features by accuracy in descending order. We measure accuracies by the validation function (details in \ref{Validation}).
        }
        \label{fig:knnindividual}
\end{figure}

In Figure \ref{fig:knnindividual} we display the results of running the KNN model on individual features; see Section \ref{setup} for experimental design details. To do this, we train our model using only a single feature in place of the entire data set. We can interpret this as projecting each data point onto the axis of that feature, so the space we are learning within is one dimensional. For each feature, we tune the model on a small batch of possible $k$ values. 

\subsubsection{Top Features} \label{top_features}

From the results of our models we now create a ranking of the most important features in our dataset for predicting GROC labels. To do this, we first define $R(m,i)$ to be the ranking of feature $i$ by model $m$. Since we have 215 features, note that $1 \le R(m,i) \le 215$ for all $i, m$. For each the metric we use to produce the ranking is the metric used to order the features in Figures \ref{fig:rindividual}, \ref{fig:svmindividual}, \ref{fig:svmsingleindividual}, \ref{fig:entropyindividual}, and \ref{fig:knnindividual}. In order to aggregate these rankings, we take the simple approach of averaging the ranking of each feature by all of our models. Let $S(i)$ be the average rank of feature $i$. Then 

\begin{equation*}
S(i) = \displaystyle \frac{1}{5}\sum_{m=1}^5 R(m,i).
\end{equation*}

\newcommand\x{11}

 This aggregates our rankings into a single score where smaller values indicate more important features.
 In Table \ref{table:topfeatures}, we show the top 30 features sorted by value.
\begin{table}[H]
\centering
\caption{Top Feature Ranks}
\csvreader[respect all,
  tabular=c c | c c,
  table head=\hline \bfseries{\hspace{\x pt}Feature\hspace{\x pt}} & \bfseries{\hspace{\x pt}Rank\hspace{\x pt}} & \bfseries{\hspace{\x pt}Feature\hspace{\x pt}} & \bfseries{\hspace{\x pt}Rank\hspace{\x pt}}\\\hline,
  late after last line=\\\hline
]{
  fig/table_4.csv
}{}{\csvlinetotablerow}
\label{table:topfeatures}
\end{table}

\subsection{Restricting to Key Features} \label{restricting}

In order to demonstrate the significance of the top 30 features (displayed in Table \ref{table:topfeatures}) that we have identified through our previous experiments as being important for predicting GROC labels, we run experiments using a dataset with only these 30 features and a dataset with all but these 30 features. We refer to the dataset of the 30 most important features as TSMLD (top subsampled MyLymeData), and we refer to the dataset of 185 remaining features as BSMLD (bottom subsampled MyLymeData).

In Table \ref{table:general_acc} we list the results for running each model on subsampled MLD (SMLD), top subsampled MLD (TSMLD), and bottom subsampled MLD (BSMLD). We list prediction accuracies for our classification models, and relevant regression values for the linear regression model. See Section \ref{setup} for experimental design details.

\begin{table}[H]
\centering
\caption{Results for running our models on subsampled MLD (SMLD), top subsampled MLD (TSMLD), and bottom subsampled MLD (BSMLD). For linear regression we list regression values and for our classificaiton models we list accuracies measured by the validation function (details in \ref{Validation}).}
\begin{tabular}{l l l l l}
  \hline \bfseries{Model} & \bfseries{SMLD} & \bfseries{TSMLD}  & \bfseries{BSMLD}\\\hline
  
  SVM & 0.484 & 0.546 & 0.436\\
  Neural Net & 0.518 & 0.506 & 0.477\\
  Decision Tree & 0.523 & 0.528 & 0.426\\
  KNN & 0.490 & 0.519 & 0.441  \\
\\
   Linear Regression \\
   \quad Adjusted $R^2$ & 0.290 & 0.274 & 0.136\\
   \quad p-value & 6.2E-36 & 6.8E-67 & 6.5E-12\\
   \hline
\end{tabular}
\label{table:general_acc}
\end{table}

By every metric, we see that the 30 features we identified (represented above by TSMLD) match or outperform the dataset of all 215 features (represented above by SMLD), suggesting that they hold a significant portion of the information which yields each model their predictive ability for the GROC label. These results also suggest possible redundancy in the dataset given the the large portion of features that contain a small amount of the information used by the models for their predictive ability. The results could also offer intuition for LymeDisease.org, the creators of this survey, about what features to focus on in designing future surveys, as well as intuition for other future survey designers. 
\section{Discussion} \label{discussion}
Here we explore and analyze the results from the previous sections.

\subsection{Predictive Significance of MyLymeData Dataset}
Our models were able to achieve a highest test accuracy of 0.613 on MLD and 0.517 on SMLD using a neural network for predicting the three classes of GROC response. We demonstrate that a significant portion of the predictive information from the dataset comes from only 30 of the 215 features. In fact, for many models our top dataset of 30 features performed better than the full dataset. 

\subsection{Antibiotics}

Of the 30 top features that we identified from the models, 20 of these features related directly to antibiotics, which suggests that many factors relating to antibiotics, including the effectiveness of antibiotic treatment (Abx\_Eff), the length of the current treatment protocol (Abx\_Dur), and the reasons why a participant is not taking antibiotic (Abx\_Not), are important predictors of a participant's GROC class.

By most of our models, Abx\_Eff was the most important feature by a large margin. This is expected because the effectiveness of the current antibiotic treatment may reflect response to antibiotic therapy generally, which GROC measures. This suggests a very close relationship between antibiotic treatment and GROC label. 
This also yields intuitive evidence that the models are 
successfully selecting the most important features, since the information offered by Abx\_Eff should make it a top feature. 

The fact that most of the features identified in the top 30 were related to the use of antibiotic treatment is important because there is currently an on-going debate about whether antibiotics are useful for treating  persistent Lyme disease. Our analysis suggests that antibiotic related questions may be the most important in predicting a patient's treatment response for those with persistent Lyme disease. This topic is explored in greater detail in a companion study analyzing the role of specific features identified in the top 30 in connection with treatment response \cite{johnson2020antibiotic}.

\subsection{Symptoms}

\begin{figure}[H]
  \centering
      \includegraphics[width=0.9\columnwidth]{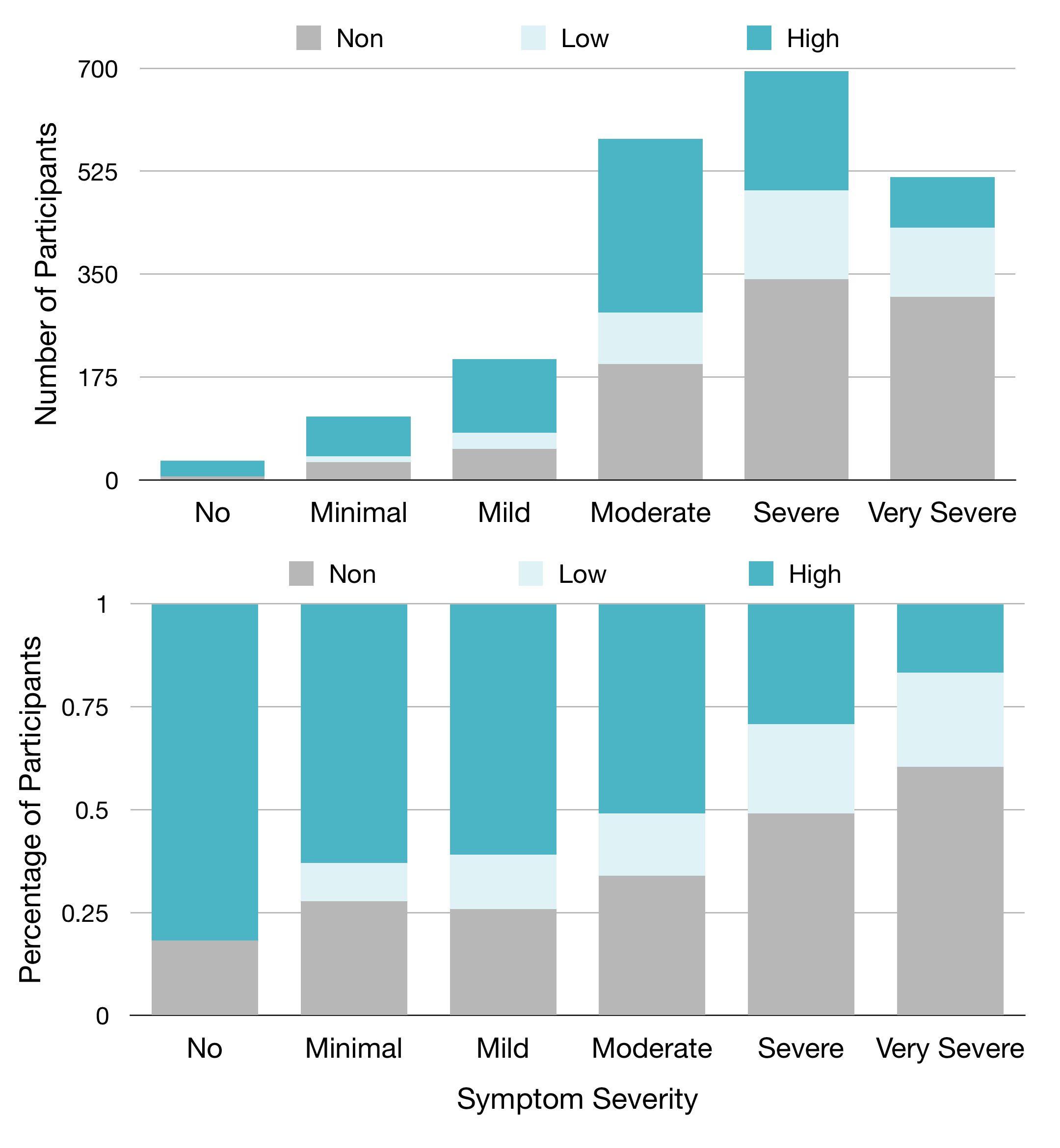}

        \caption{
                (Top) A stacked bar graph with participants from each GROC class by their answers to severity of fatigue symptoms (Sx\_Sev\_1), increasing in severity. (Bottom) a normalized version of the bar graph that gives the percentage representation of each GROC class by response to Sx\_Sev\_1, for better visualization of trends.
        }
        \label{fig:fatigue}
\end{figure}

Of the top 30 features we identified from our models, nine of these features are from the 13 named Sx\_Sev\_i that ask about the current severity of specific symptoms, suggesting that these symptoms or the severity of these symptoms are important predictors of GROC class. 
Based on our ranking metric, the second most important predictor of GROC label is feature Sx\_Sev\_1, which asks the current severity of fatigue symptoms. To visualize this relationship, in Figure \ref{fig:fatigue} we provide a chart relating participants' responses to Sx\_Sev\_1 with GROC label.

Figure \ref{fig:fatigue} demonstrates a clear relationship between Fatigue severity and GROC treatment response label, as it shows that participants who report no minimal, or mild symptom severity for fatigue are most often high responders, while those with severe or very severe fatigue are most often non-responders. This suggests that the severity of fatigue symptoms could be a useful metric in determining GROC for Lyme disease patients, which matches this result in our experiment. 

\subsection{Branching}
Although feature Abx did not appear in our list of top 30 features, a more comprehensive analysis reveals the true importance of this feature. Feature Abx identifies whether the participant is taking antibiotics, and this feature is used for branching purposes, so features concerning specifics of antibiotic treatment are only asked to patients who indicated in this feature that they are currently taking antibiotics. Thus, any feature that relies on the branching effect of Abx will inherently contain all the information contained in Abx, since for any such feature all patients who are not taking antibiotics would be grouped together. For this reason, Abx does not serve the purpose that the top 30 features list represents; our intention was to find 30 features that effectively predict GROC labels, and since other features in this list that result from the branching of Abx contain all the relevant information about whether a patient is taking antibiotics, Abx would not increase the predictive ability of models run on this set and or offer unique information. We can see this represented in our analysis; Abx had an average rank of 31 amongst all metrics other than SVM drop-out, in which it ranked 191 0f 215. Since the SVM drop-out was the only metric that measured a feature by examining the loss from removing it, these results can be explained by the lack of unique information offered by Abx as explained above while also demonstrating the importance of Abx independent of the rest of the dataset. 

The branching structure also affects the importance and interpretability of the top feature set by 
yielding some features with substantial predictive information unrelated to the purpose of the question being asked. Within the top feature set, those that appear to be most affected by this are the Abx\_i features and Abx\_IM\_6. Upon further inspection, these features have little significance aside from maintaining the branching information from Abx that identifies which participants are currently taking antibiotics, yet the little information they add after this split make them more important than Abx based on information content alone. On one hand, this indicates that the subject matter of these features may not be as important as the analysis initially suggests. Yet, this does not take away from the purpose of these features within the top features list, which is to hold as much significant information as possible in a small subset of the features, since these features do hold important information from the branching in Abx. 

\subsection{Machine Learning for Survey Data}
Here we outline two important factors one should consider when applying machine learning methods to a survey dataset. 

One consideration comes in choosing the models and metrics of evaluation. Within the goal of finding important features in a data set, one can either evaluate the predictive information from a single feature, or remove a single feature from the data set and measure the effects on performance, as in an abalation study. The former metric measures the predictive information from a single feature, while the later metric measures the amount of information unique to the specific features in comparison to the other features. Of our five metrics for measuring importance of individual features used in our analysis, four measure features independantly of the data set, and one, the SVM dropout metric, removed a feature from the data set to measure performance. Our results align with this intuition of the metrics; we explored this distinction of metrics in the previous section in the context of feature Abx.

When applying machine learning models to survey data for feature selection, it is also important to contextualize the data and results within the branching structure, which can affect the meaning and interpretation of the results. We describe the effects of the branching structure on our analysis in the previous section.

\section{Conclusion}

We provide results of applying various simple feature selection techniques to the LDo MLD dataset.  These techniques provide insights to which participant features are most important in determining participant GROC responder status.  These insights may be valuable to medical professionals in determining the factors that are most predictive of treatment response.

Furthermore, these results demonstrate the potential and efficacy of these simple feature selection techniques for determining important aspects of datasets.  We expect that similar experiments could be valuable in survey development, for survey design and reduction of survey fatigue, as well as in other areas of science.

\paragraph{Conflict of Interest}

The authors declare no conflict of interest.

\paragraph{Acknowledgment}

The authors are grateful to and were partially supported by NSF CAREER DMS \#1348721 and NSF BIGDATA DMS \#1740325.  The authors would like to thank LymeDisease.org for the use of data derived from the MyLymeData patient registry, Phase 1 27 April 2017. The authors thank the patients for their contributions to MyLymeData. We also thank Mira Shapiro for her advice and expertise.

\footnotesize{

\bibliography{main}
\bibliographystyle{unsrt}

\noindent
} 
\vspace{5pt}
\onecolumn
\appendix
\section*{Appendix}
\begin{center}
\captionof{table}{Descriptions of relevant features. This includes all features mentioned in the paper. The Feature Name is the name used throughout the paper and the Variable Name is the original name used by the MyLymeData dataset.}
\vspace{10pt}
\resizebox{0.80\textwidth}{!} {
\csvreader[respect all,
  tabular=l l l,
  table head=\hline\bfseries{Feature Name} & \bfseries{Variable Name} & \bfseries{Description} \\\hline,
  late after last line=\\\hline
]{
  fig/table_6.csv
}{}{\csvlinetotablerow}
}
\label{table:attachment}
\end{center}

\end{document}